\input{aipcheck}

\documentclass[
    ,final            
  ]
  {aipproc}

\layoutstyle{6x9}

\begin{document}

\title{Timelike small x Resummation for Fragmentation Functions 
}


\author{S.\ Albino}{
  address={{II.} Institut f\"ur Theoretische Physik, Universit\"at Hamburg,\\
             Luruper Chaussee 149, 22761 Hamburg, Germany}
}

\author{P.\ Bolzoni}{
  address={{II.} Institut f\"ur Theoretische Physik, Universit\"at Hamburg,\\
             Luruper Chaussee 149, 22761 Hamburg, Germany}
}

\author{B.\ A.\ Kniehl}{
  address={{II.} Institut f\"ur Theoretische Physik, Universit\"at Hamburg,\\
             Luruper Chaussee 149, 22761 Hamburg, Germany}
}

\author{A.\ Kotikov}{
  address={{II.} Institut f\"ur Theoretische Physik, Universit\"at Hamburg,\\
             Luruper Chaussee 149, 22761 Hamburg, Germany}
  ,altaddress={Bogoliubov Laboratory of Theoretical Physics, JINR, 141980 Dubna, Russia} 
}

\begin{abstract}
The status of small x resummation in the timelike kinematics is discussed.
We present a general procedure to extract the large logarithms of x in the 
$\overline{\rm MS}$ factorization scheme and to resum them in a closed form. 
New results for the doubly-logarithm-resummed 
coefficient functions will be reviewed.
All our resummation formulae are in 
agrement with the fixed NNLO computations recently done by other groups in 
the $\overline{\rm MS}$ scheme.
\end{abstract}

\maketitle


To fix the ideas,
we consider the cross section for the semi-inclusive hadron 
production in electron-positron annihilation:
\begin{equation}
e^+(k_1) + e^-(k_2) \rightarrow V^*(q)\rightarrow h(p_h)+X,
\label{epemg}
\end{equation}
where $V^*$ is a virtual vector boson with virtuality $Q^2=q^2=(k_1+k_2)^2$ and $X$
stands for any allowed hadronic final state.
Here we are interested in the differential cross section for the single hadron production
 $d\sigma^h(x,Q^2)/dx$
where $x$ is the scaled momentum fraction of the produced hadron $h$:
\begin{equation}
x=\frac{2p_h\cdot q}{Q^2},\qquad 0\leq x \leq 1.
\end{equation} 
According to QCD facotrization the cross section can be written as 
the convolution of the partonic cross section to produce a parton $k$
with scaled momentum fraction $z=x/z'$ with the fragmentation function
$D_{k/h}(z')$ from the parton the hadron $h$
with scaled momentum fraction $x$ (see e.g. Ref.\cite{Albino:2008gy}):
\begin{equation}
\frac{d\sigma^h}{dx}(x,Q^2)=\sum_k \int_x^1 \frac{dz'}{z'}\frac{d\hat{\sigma}_{e^+e^-\rightarrow k}}{dz}
D_{k/h}(z'). 
\end{equation}
Usually the partonic cross section $d\hat{\sigma}_{e^+e^-\rightarrow k}/dx$ is written
in terms of the coefficient functions $C_k(x,Q^2)$ defined as
\begin{equation}
C_k(x,Q^2)=\frac{1}{\sigma^{(\rm ew)}_{k}}\,\frac{d\hat{\sigma}_{e^+e^-\rightarrow k}}{dx},
\label{defcoeff}
\end{equation}
where $\sigma^{(EW)}_{k}$ contains all the electro-weak over-all factors.

Perturbation theory fails when the fraction $x$ of available energy carried away by the observed particle is 
too low, because large logarithms spoil the convergence of the perturbative series.
The largest logarithms, the {\it double logarithms} (DLs), in the splitting functions that determine the evolution of the
fragmentation functions have been computed to all orders a long time ago \cite{Mueller:1981ex}, and have even been used to perform 
 LO global fits in QCD  
\cite{Albino:2005gd,Albino:2005gg} to data measured at the smallest $x$ values.
The DLs appearing in the coefficient functions $C_k(x,Q^2)$
 are not resummed at LO and are expected to be not 
as important as those appearing in the evolution because they only appear at and 
beyond NLO in QCD.
However, the inclusion of the DLs in the coefficient functions could make a 
significant improvement to the 
accuracy of cross section calculations making the analysis of Refs.\ \cite{Albino:2005gd,Albino:2005gg}
feasible also at NLO.
The complete DL contribution to partonic cross sections has been calculated 
in Ref.\cite{Mueller:1982cq} for the case in which the collinear singularities are regularized by giving a 
small mass $m_g$ to the gluon, the so-called massive gluon (MG) regularization scheme. 
The inconsistency noted in \cite{Albino:2008gy,Albino:2011si} between the NNLO DLs calculated 
from the resummed result in Ref.\ \cite{Mueller:1982cq} and 
those calculated from the fixed order result in Refs.\ \cite{Rijken:1996vr,Rijken:1996ns,Rijken:1996npa,Mitov:2006ic,Mitov:2006wy,Blumlein:2006rr}
is not surprising because the two computations
were carried out in two different regularization and factorization schemes, namely the MG and the 
$\overline{\rm MS}$ scheme.  
The DLs of the gluon coefficient functions have been computed for the first time in the 
$\overline{\rm MS}$ scheme very recently by us in \cite{Albino:2011si}. 
Here we will rederive the same result in a less formal but more simple way.


Our goal is to get a  
resummed cross section in the $\overline{\rm MS}$ scheme for which the dimensional regularization
is necessary.
To extract the leading logarithmic behavior, we exploit the factorization of the 
single gluon probability emission in the soft-collinear limit. 
This is a consequence of the eikonal approximation
and color coherence as it has been proven a long time ago in 
\cite{Bassetto:1982ma,Bassetto:1984ik}. This result can be used   
\cite{Albino:2011si} to 
obtain the probability gluon emission in $d=4-2\epsilon$ dimensions:  
\begin{equation}
dw(x,z,\epsilon)=2 \mathcal{C}_i \, a_s \left(\frac{\mu^2}{Q^2}\right)^\epsilon \frac{(4\pi)^\epsilon}{\Gamma(1-\epsilon)} 
\frac{dx}{x^{1+2\epsilon}}\, \frac{dz}{z^{1+\epsilon}},
\label{probemission}
\end{equation} 
where $a_s=\alpha_s/2 \pi$, $\mu$ is the dimensional regularization scale taken here and in the following
equal to the renormalization scale and where 
$z=(1-\cos\theta)/2$ with $\theta$ the scattering angles of the emitted soft gluon with respect to
the hard jet direction. Here $\mathcal{C}_i=C_A$ for a gluon jet and $\mathcal{C}_i=C_F$ for a quark jet.
The expression for the probability emission given in Eq.(\ref{probemission}) is 
what we need to obtain the gluon probability density in dimensional regularization. 
\begin{figure}
  \includegraphics[height=.37\textheight]{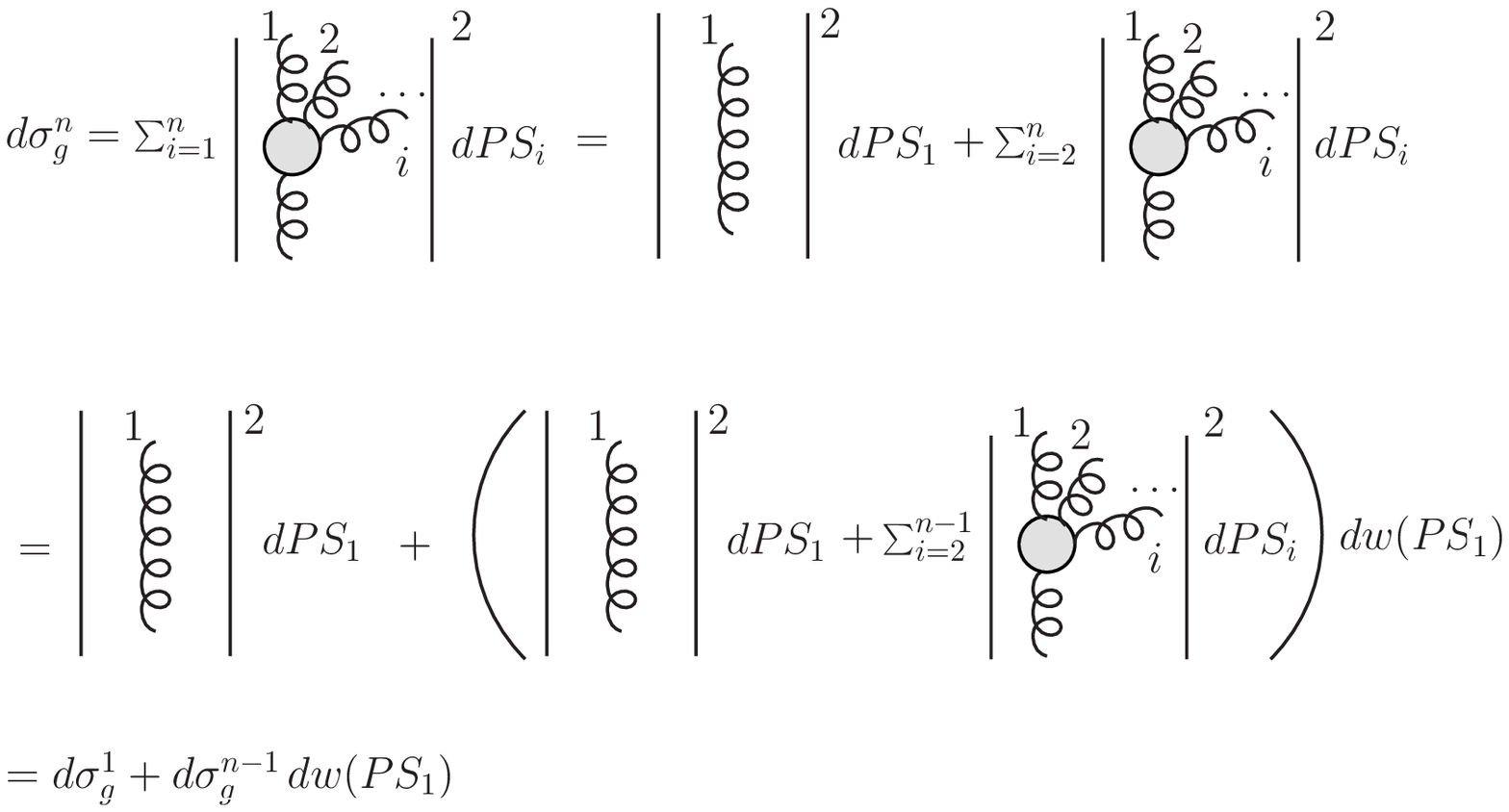}
  \caption{Here $d\sigma_g^n$ represents the cross section for the production 
  of a gluon up to real corrections of order $n$, $dPS_i$ is the $i$-particles
  phase space and $dw(PS_1)=dw(x,z,\epsilon)$ after that azimutal integration
  is performed. The factorization of the single gluon probability
  emission has been used in the third step.}
\label{heuristic}
\end{figure}
Fig.\ref{heuristic} shows a diagrammatic derivation of a consistency relation
for the differencial cross section for gluon jet production which is:
\begin{equation}
d\sigma_g^n=d\sigma_g^1+d\sigma_g^{n-1}dw(x,z,\epsilon).
\end{equation}
Now introducing $G(x,\epsilon)$ the gluon distribution density and
taking the limit $n\rightarrow \infty$, we obtain immediately
the following bootstrap equation for it:
\begin{equation}
x^{1+2\epsilon}\mathcal{G}(x,z,\epsilon)=
\delta(1-x)+\int_x^1 dx' \int_z^1 dz'\,K(x',z',\epsilon)\, x'^{1+2\epsilon} \mathcal{G}(x',z',\epsilon), 
\label{mastereq}
\end{equation}
where $K(x,z,\epsilon)=dw(x,z,\epsilon)/dx\,dz$ and where the factor of 
$x^{1+2\epsilon}$ represents our normalization
coming from the explicit computation for the first gluon emission with $n=2$.
In Eq.(\ref{mastereq}) the value $\mathcal{G}(x,z=0,\epsilon)=G(x,\epsilon)$
should be taken only at the end of the computation. This ensures the strong angular
ordering of the emitted gluons, which is necessary to extract correctly 
the leading logarithms as proven in Refs.\cite{Mueller:1981ex,Bassetto:1982ma}.
Performing the Mellin transform,
\begin{equation}
f(\omega)=\int_0^1\, dx x^\omega f(x); \qquad \omega=N-1,
\end{equation}
of Eq.(\ref{mastereq}), then performing also the $z$-integrals,  
solving recursevely for $\mathcal{G}(x,z,\epsilon)$ and finally putting $z=0$
we obtain:
\begin{equation}
G(\omega,\epsilon)=1+\sum_{k=1}^\infty \left[2a_sC_A\frac{(4\pi)^\epsilon}{\Gamma(1-\epsilon)}
\left(\frac{\mu^2}{Q^2}\right)^\epsilon\right]^k\frac{(-1)^k}{k!\epsilon^k}\prod_{l=1}^k
\frac{1}{\omega-2l\epsilon}.
\label{solution}
\end{equation}
According to the QCD factorization theorem we have that all the collinear singularities in 
Eq.(\ref{solution}) should be factorized. In the $\overline{\rm MS}$ factorization
scheme this is done requiring Eq.(\ref{solution}) to be compared with \cite{Curci:1980uw,Catani:1994sq},
\begin{equation}
G(\omega,\epsilon)=G^{\overline{\rm MS}}\left(\omega,a_s,\frac{Q^2}{\mu_F^2}\right)
\exp\left[-\frac{1}{\epsilon}\int_0^{a_s(\mu^2/\mu_F^2)^\epsilon S_\epsilon}\frac{da}{a}
\gamma^{\overline{\rm MS}}(\omega,a)\right],
\label{factorization}
\end{equation}
where $S_\epsilon=(4\pi)^\epsilon e^{-\epsilon\gamma_E}$ with $\gamma_E$ the Euler number
and where $\mu_F$ is the arbitrary factorization
scale. The direct comparison of the two Eqs.(\ref{solution},\ref{factorization}) is non
trivial. As shown in \cite{Albino:2011si} a possible way to do this is to compare a simple 
differential equation satisfied by $G(\omega,\epsilon)$ given in Eq.(\ref{solution}) 
with the same differential equation this time 
obtained by use of the factoriation constraint given in Eq.(\ref{factorization}). 
We report here the result which is given by:
\begin{equation}
G^{\overline{\rm MS}}\left(\omega,a_s,\frac{Q^2}{\mu_F^2}\right)=C^{\overline{\rm MS}}(\omega,a_s)
\exp\left[\gamma^{\overline{\rm MS}}(\omega,a_s)\log\left(\frac{Q^2}{\mu_F^2}\right)\right],
\label{result}
\end{equation}
where
\begin{equation}
\gamma^{\overline{\rm MS}}(\omega,a_s)=\frac{1}{4}\left[-\omega+\sqrt{\omega^2+16C_Aa_s]}\right];
\quad C^{\overline{\rm MS}}(\omega,a_s)=\left[\frac{\omega}{4\gamma^{\overline{\rm MS}}(\omega,a_s)
+\omega}\right]^{\frac{1}{2}}.
\end{equation}
Expanding the result in Eq.(\ref{result}) up to NNLO we obtain perfect agreement with the leading logarithmic
terms of the fixed order result computed in the literature in the same scheme (see e.g. Eqs.(A.3,A.6) in Ref.\cite{Blumlein:2006rr}).

We conclude noting that attaching the gluon jet to the quark lines of the LO process in Eq.(\ref{epemg}), 
we have that the gluon coefficient function
is according to Eq.(\ref{defcoeff}):
\begin{equation}
C^{\overline{\rm MS}}_g(\omega,as)=\frac{2C_F}{C_A}\left[C^{\overline{\rm MS}}(\omega,a_s)-1\right].
\label{finalresult}
\end{equation}
This result enables us to resum all the DLs in the gluon coefficient function in the $\overline{\rm MS}$
factorization scheme for the first time and according to 
Refs.\cite{Altarelli:1979kv,Baier:1979sp} to have full control
over all the large logarithms in all the coefficient functions at NLO. Our result Eq.(\ref{finalresult})
is also a key ingredient in the determination of the {\it single logarithms} in the timelike splitting
functions, which is left to a forthcoming paper \cite{Albinofuture}.




\bibliographystyle{aipproc}   

\bibliography{proc_Bolzoni}

\IfFileExists{\jobname.bbl}{}
 {\typeout{}
  \typeout{******************************************}
  \typeout{** Please run "bibtex \jobname" to optain}
  \typeout{** the bibliography and then re-run LaTeX}
  \typeout{** twice to fix the references!}
  \typeout{******************************************}
  \typeout{}
 }

\end{document}